\documentclass[conference]{IEEEtran}
\usepackage{cite}
\usepackage{amsmath,amssymb,amsfonts}
\usepackage{algorithmic}
\usepackage{graphicx}
\usepackage{textcomp}
\usepackage[table,svgnames]{xcolor}
\usepackage{listings}
\usepackage{hyperref}

\hypersetup{
    pdftitle={Parameterization-Independent Importance Sampling of Environment Maps},
    pdfsubject={Technical Report},
    pdfauthor={Martin Lambers}
  }

\lstdefinelanguage[OpenGL]{C}[ANSI]{C}{%
    morekeywords={bool,bvec2,bvec3,bvec4,ivec2,ivec3,ivec4,vec2,vec3,vec4}
}
\definecolor{cgblue}{rgb}{0.2,0.2,0.7}
\begin{document}

\title{Parameterization-Independent Importance Sampling of Environment Maps}

\author{\IEEEauthorblockN{Martin Lambers}
\IEEEauthorblockA{\textit{Computer Graphics Group} \\
\textit{University of Siegen}\\
Siegen, Germany \\
martin.lambers@uni-siegen.de}
}

\maketitle

\begin{abstract}
   Environment maps with high dynamic range lighting, such as daylight sky maps, require
   importance sampling to keep the balance between noise and number of samples
   per pixel manageable.
   Typically, importance sampling schemes for environment maps are based
   directly on the map parameterization, e.g.~equirectangular maps, and do not
   work with alternative parameterizations that might provide better sampling
   quality.
   In this paper, an importance sampling scheme based on an equal-area
   projection of the sphere is proposed that is easy to implement and works
   independently of the environment map parameterization or resolution.
   This allows to apply the same scheme to equirectangular maps, cube map
   variants, or any other map representation, and to adapt the importance
   sampling granularity to the requirements of the map contents.
\end{abstract}


\section{Introduction}

In physically based rendering, high dynamic range environment maps can provide
realistic lighting of scenes. In such maps, the lighting is often distributed
very unevenly. For example, in daylight sky maps, the small region that
represents the sun is thousands of times brighter than other regions of the sky.
Importance sampling is therefore required to keep noise in the
rendering result at acceptable levels while limiting the required number of samples per pixel.

The importance sampling scheme for environment maps is typically based directly
on the environment map parameterization. For example, the Mitsuba and pbrt
renderers support environment maps using the equirectangular
(latitude-longitude) parameterization and build auxiliary maps in the same
parameterization to create light direction samples and compute corresponding
probability density function values~\cite{pharr2016pbr}. However,
equirectangular maps are known to suffer from strong distortions especially in
the polar areas, which reduces sampling quality. Other parameterizations of
environment maps, such as classical cube maps~\cite{greene1986cubemap} or low-distortion
variants thereof~\cite{lambers2020cubemaps} provide better sampling quality, but supporting them
requires tailored importance sampling schemes.

We propose to decouple the importance sampling scheme from the
environment map parameterization and its resolution. Our importance sampling scheme is based on an
equal-area projection between sphere and square, which simplifies 
the required computations. It can be applied to all environment map parameterizations
and allows adapting its granularity to reduce computational costs.

\section{Related Work}

This paper focusses on importance sampling and multiple importance sampling
for Monte Carlo renderers, not on alternative techniques like product importance
sampling~\cite{clarberg2008product,estevez2018product} or optimizations for
special cases~\cite{kollig2002diffuse}, methods that focus
on visibility considerations~\cite{brogers2014importance,bitterli2015portal}, or
methods for dynamic environment maps~\cite{havran2005video}.

Importance sampling strategies for environment maps
in Monte Carlo renderers typically compute importance
maps based on the environment map parameterization, and use them as the base
to generate light direction samples and to compute the underlying probability
density function. For example, both the Mitsuba and pbrt renderers support
equirectangular environment maps and precompute equirectangular importance
maps to enable importance sampling~\cite{pharr2016pbr}. Agarwal et al.~apply
hierarchical structuring to
equirectangular maps~\cite{agarwal2003structured}, and Lu et al.~base importance maps on the cube map
representation used in their real-time renderer~\cite{lu2013realtime}.

In these cases, supporting different types of environment map parameterizations is difficult
since they require derivation of tailored importance sampling schemes.
This paper therefore proposes to decouple the importance sampling from the environment
map parameterization.

\section{Method}

A basic environment map implementation in a renderer typically only needs one
function $L(\vec{d})$ to return a radiance spectrum or RGB sample for a lookup direction
$\vec{d}$. 

To support importance
sampling, an implementation additionally needs to a way to generate light
direction samples according to a suitable probability density function. To
support multiple importance sampling, this functionality is separated into a
function $p(\vec{d})$ that computes the probability density function value for a
given direction and a function $d()$ that generates a random light direction sample
according to $p$.

Traditionally, these functions all work on the same environment map
parameterization. In our approach, the sampling function
$L(\vec{d})$ still uses the original environment map representation to make sure
that sampling quality remains the same, while functions $p(\vec{d})$ and $d()$ work on
importance maps that are tailored to their needs.
This means that these functions only need to be implemented once and will work
with any environment map parameterization.  A minimal C++ interface that
implements this approach is shown in Fig.~\ref{fig:envmap-interface}.

\begin{figure}[t]
\begin{lstlisting}
class EnvMap {
  // Functions implemented in subclasses such as
  // EnvMapEquirect, EnvMapCube:

  // Sample the environment map.
  virtual Spectrum L(vec3 d);

  // Functions implemented only once, independent
  // of subclasses:

  // Return the pdf value for direction d:
  float p(vec3 d);
  // Generate a random direction according to p:
  vec3 d();
};
\end{lstlisting}
\caption{\label{fig:envmap-interface}
Minimal interface of an environment map that supports multiple importance sampling.}
\end{figure}

For good importance sampling results, the value of the probability density
function $p$ for a direction $\vec{d}$ must be proportional to a suitable importance
measure $I(\vec{d})$ for that direction: $p(\vec{d}) \propto I(\vec{d})$.
Here the sum of radiances over the spectrum is used as the importance measure:
$I(\vec{d}) = \sum_{i=1}^{k} L(\vec{d})$ for a spectrum with $k$ samples ($k=3$
for RGB environment maps). Note that other measures can be used, e.g.~a
luminance value computed from the spectrum as used by Mitsuba and pbrt.

Furthermore, to be a valid probability distribution function, $p$ must integrate
to 1 over the surface of the unit sphere: $\int_{S^2} p d\omega = 1$.

Our implementation of the functions $d()$ and $p(\vec{d})$ is based on an importance
map $M$, a sorted importance map $M_s$, and a cumulative sorted importance map
$M_{cs}$.
These maps are based on an equal-area projection between sphere and square
described in Sec.~\ref{sec:proj}.
Their construction is described in Sec.~\ref{sec:maps}.
Values of the probability density function $p(\vec{d})$ can be taken directly from 
map $M$, as described in Sec.~\ref{sec:p}, and light direction samples for
function $d()$ can be generated based on maps $M_{cs}$ and $M_s$, as described
in Sec.~\ref{sec:d}.

\subsection{Equal Area Projection between sphere and square}
\label{sec:proj}

Using an equal area projection between sphere and square to generate the
importance maps has the advantage that uniformly distributed random numbers on
the 2D square translate directly to uniformly generated random numbers on the
unit sphere surface, which simplifies both the sampling of light directions and
the evaluation of the probability density function.

The unit sphere surface is first projected to the unit disk using the Lambert
Azimuthal Equal-Area projection~\cite{snyder1987mapprojections} and the unit
disk is then projected to the unit square using Shirley's equal area
projection~\cite{shirley1997map}.

The resulting mapping function $m: S^2\rightarrow [0,1]^2$ first maps
a direction $\vec{d}$ represented as latitude $\theta$ and longitude $\varphi$
to a point on the unit disk represented as radius $r$ and angle $\alpha$, and
then to a point $(u,v)$ in the unit square:\\
$r = \sin\left(\frac{\frac{\pi}{2} - \theta}{2}\right) \quad
\alpha = \varphi - \frac{\pi}{2} \quad
\alpha' = 
\begin{cases}
\alpha          & \alpha \geq -\frac{\pi}{4}\\
\alpha + 2\pi   & \alpha <    -\frac{\pi}{4}
\end{cases}$\\
$(u',v') = 
\begin{cases}
(r,\frac{\alpha' r}{\frac{\pi}{4}})                       & \alpha' < \frac{\pi}{4}\\
(-(\alpha' - \frac{\pi}{2}) \frac{r}{\frac{\pi}{4}},r)    & \alpha' < \frac{3\pi}{4}\\
(-r,(\alpha' - \pi) \frac{-r}{\frac{\pi}{4}})             & \alpha' < \frac{5\pi}{4}\\
(-(\alpha' - \frac{3\pi}{2}) \frac{-r}{\frac{\pi}{4}},-r) & \text{otherwise}\\
\end{cases}
$\\
$u=\frac{1}{2} (u'+1) \quad v=\frac{1}{2} (v'+1)$

The inverse mapping is defined as follows:\\
$u'=2u-1\quad v'=2v-1$\\
$(r,\alpha) =
\begin{cases}
(u',\frac{\pi}{4}\frac{v'}{u'})       & u'^2 > v'^2\\
(v',\frac{\pi}{2}-\frac{\pi}{4}\frac{u'}{v'})      & u'^2 \leq v'^2\\
(0,0)                                           & u' = v' = 0\\
\end{cases}
$\\
$\theta = \frac{\pi}{2} - 2 \sin^{-1}(r) \quad
\varphi = \alpha + \frac{\pi}{2}$\\

Note that this equal-area map is only used to generate light directions and
to look up probability density function values. The sampling of the
environment still takes place in the original environment map parameterization.
This means that we benefit from the equal-area property of the map, which results in
significant simplifications, but are unaffected by its distortions and
discontinuities~\cite{lambers2016mappings}.

\begin{figure}[t]
\includegraphics[width=.48\linewidth]{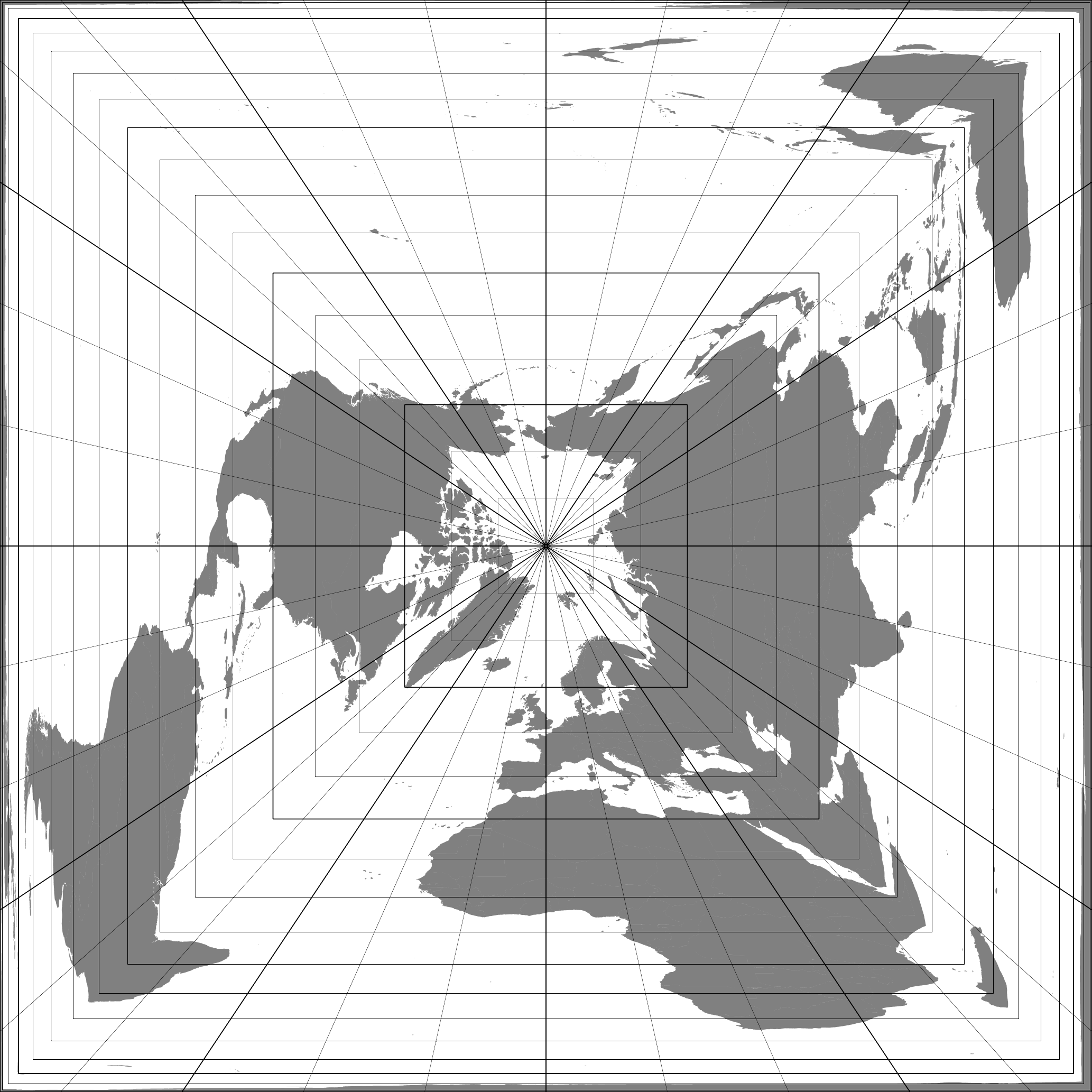}\hfill
\includegraphics[width=.48\linewidth]{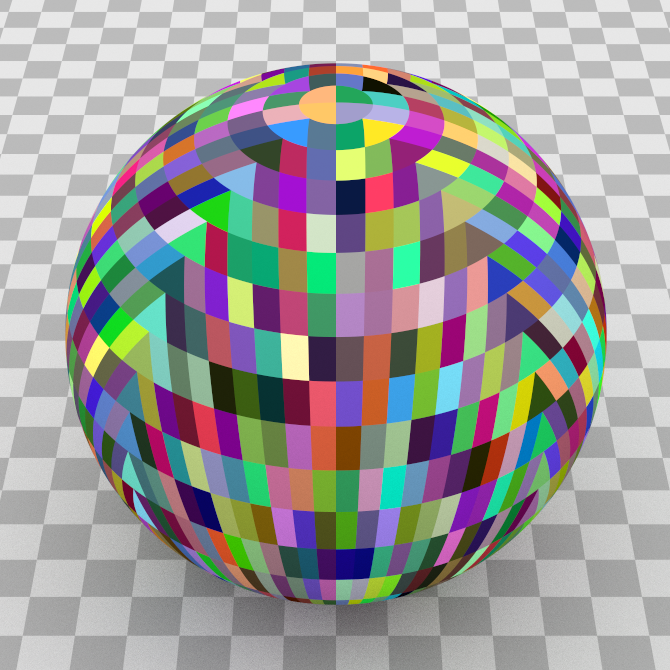}
\caption{\label{fig:map-example}
The equal area square map of the sphere. Left: the Earth's land masses
on the square map; the south pole is mapped to the border. Right: a colored
raster on the square map reprojected onto the sphere.}
\end{figure}

The map projection is demonstrated in Fig.~\ref{fig:map-example} using the
Earth's land masses as an illustration for the mapping from sphere to square.
For the inverse mapping, the square map is divided into a raster of $32\times
32$ colored squares and reprojected onto a sphere. Angular
distortions and discontinuities are visible, but all reprojected areas are of the same size.

\subsection{Importance Maps}
\label{sec:maps}

Three maps of size $N \times N$ are precomputed using the map projection defined
in Sec.~\ref{sec:proj} as a base for the importance sampling scheme:
\begin{enumerate}
\item The importance map $M$ which in each bin $(i,j)$
stores the importance measure for the corresponding direction $\vec{d}$,
normalized so that the sum over all $N \times N$ bins is 1.
First, importance values $M[i][j]=
I\left(m^{-1}\left(\frac{i+\frac{1}{2}}{N}, \frac{j+\frac{1}{2}}{N}\right)\right)$ are computed and
from them the total importance sum
$I_{\text{total}}=\sum_{i=1}^{N}\sum_{j=1}^{N}M[i][j]$. Afterwards, each bin
value is divided by $I_{\text{total}}$.
\item The sorted importance map $M_s$ which stores bin indices so that
$M[M_s[i][j]]$ is sorted according to descending normalized importance.
In practice, $M_s$ can simply store 1D integer indices $\{0, \ldots, N^2-1\}$.
\item The cumulative sorted importance map $M_{cs}$ which stores cumulative
normalized importances according to $M_s$ so that $M_{cs}[0][0] = \max_{i,j} M[i][j]$
and $M_{cs}[N-1][N-1] = \sum_{i=1}^{N}\sum_{j=1}^{N} M[i][j]=1$.
\end{enumerate}

\begin{figure}[t]
\includegraphics[height=.29\linewidth]{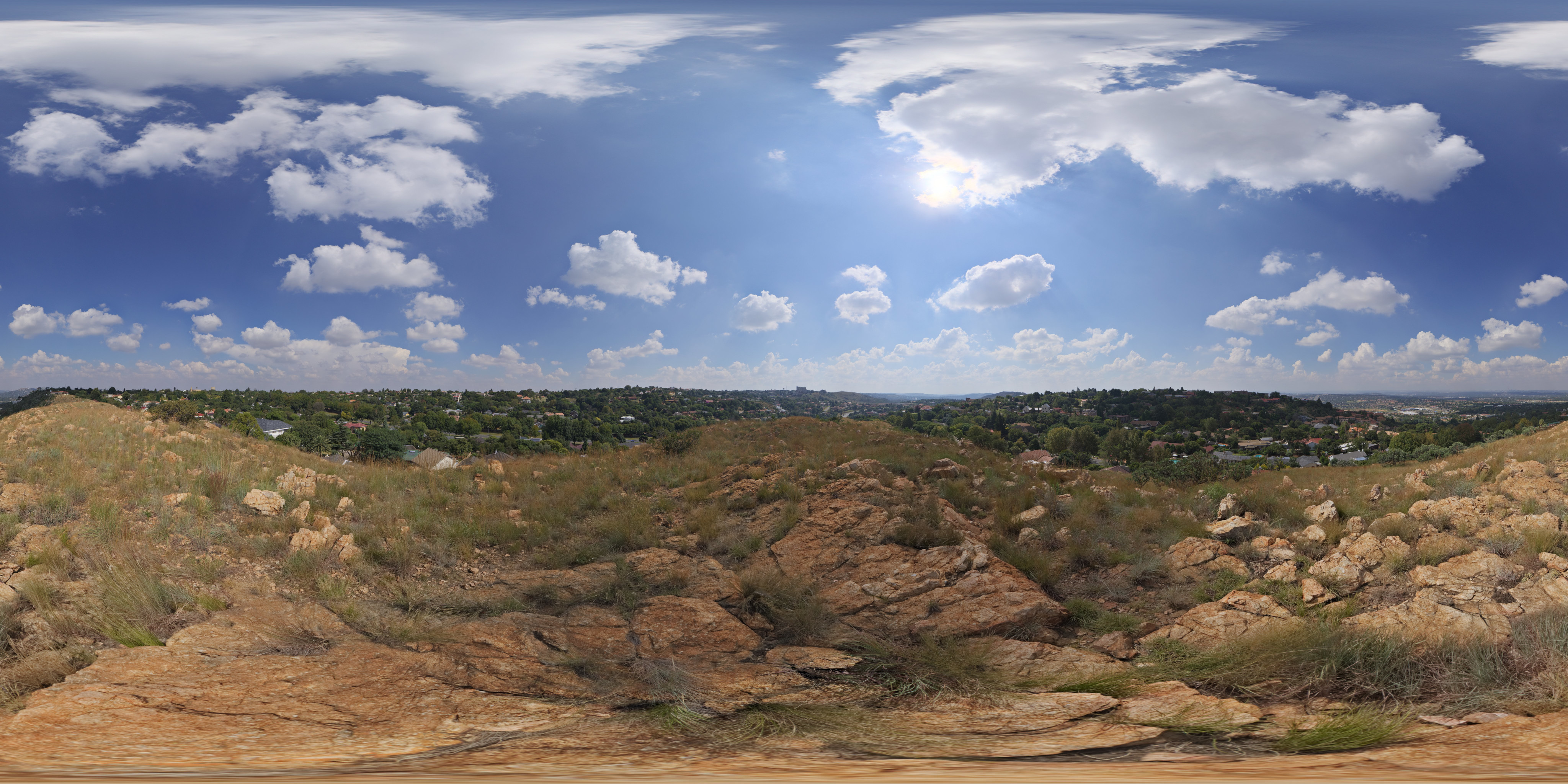}\hfill
\includegraphics[height=.29\linewidth]{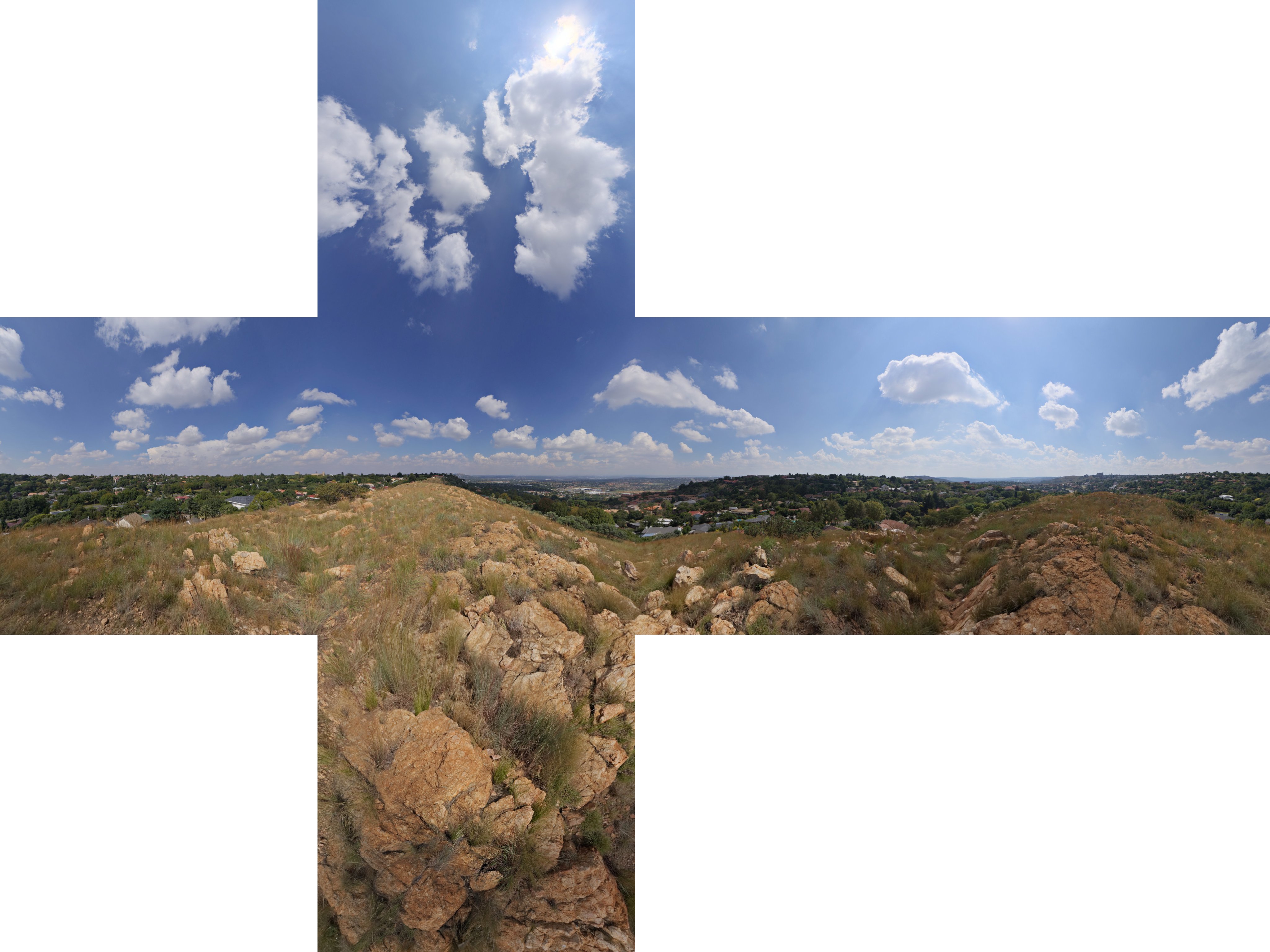}
\caption{\label{fig:envmap}
A daylight sky environment map in equirectangular (left) and cube map (right)
parameterization.}
\end{figure}

\begin{figure}[t]
\includegraphics[width=.48\linewidth]{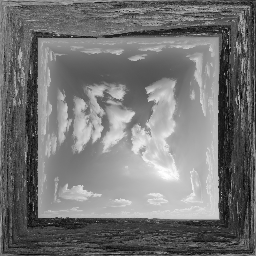}\hfill
\includegraphics[width=.48\linewidth]{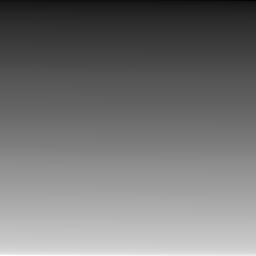}\hfill
\caption{\label{fig:impmaps}
The importance maps $M$ (left) and $M_s$ (right) computed from
the environment map shown in Fig.~\ref{fig:envmap}.}
\end{figure}

The map size $N\times N$ is independent of the resolution of the environment map.
This approach assumes approximately constant importance of all directions covered by one bin
reprojected onto the sphere. Therefore, $N$ should be chosen so that the
dominant light sources in the environment map can be sampled with enough precision,
i.e.~the bin size is small enough to provide a reasonable representation of them.
Only extreme cases of very bright light
sources covering only a few pixels in the environment map parameterization
require to chose
$N$ so that the total number of importance map entries is roughly equal to the
resolution of the original environment map; in most
cases, $N$ can be chosen significantly lower to save memory and computation
time.

Fig.~\ref{fig:envmap} shows a daylight sky environment map in equirectangular
and cube map parameterizations. Fig.~\ref{fig:impmaps} shows the maps $M$ and
$M_s$ computed from either of those parameterizations for $N=256$.

\subsection{Probability density function}
\label{sec:p}

By construction, the entries in map $M$ represent the importance of
the corresponding directions in the environment map. To compute a
probability density function value from them, the
normalized map entries need to be divided by the size $s$ of one bin on the unit sphere:
$s=\frac{4\pi}{N^2}$. This enforces the requirement that $\int_{S^2} p d\omega = 1$.

To get the probability density function value for a given direction $\vec{d}$,
map coordinates $(u,v) = m(\vec{d})$ are translated into a bin index $(i,j)$:\\
$(u,v)=m(\vec{d}) \quad (i,j)=(\lfloor Nu \rfloor, \lfloor Nv \rfloor) \quad
p(\vec{d}) = \frac{M[i][j]}{s}$

\subsection{Sampling light directions}
\label{sec:d}

To generate light direction samples with the probability density function $p$
described in Sec.~\ref{sec:p}, maps $M_{cs}$ and $M_s$ are used as follows:
\begin{itemize}
\item Generate a uniformly distributed random number $r \in [0,1)$
\item Find the entry $(i',j')$ with the cumulative normalized importance $r$ in the map
$M_{cs}$. Binary search can be applied for this purpose.
\item Get the corresponding bin index in map $M$ from $M_{s}$: $(i,j)=M_s[i'][j']$.
\item Based on uniformly distributed random numbers $s_0, s_1 \in [0,1)$, create
map coordinates $(u,v)$ inside the bin $(i,j)$:\\
$u = \frac{i + s_0}{N} \quad v = \frac{j + s_1}{N}$
\item The direction sample is $\vec{d} = m^{-1}(u,v)$.
\end{itemize}

Uniform sampling inside a bin works because the importance map projection
is equal area, otherwise the samples would not correspond to uniformly
distributed samples on the unit sphere surface patch represented by the bin.

\begin{figure}[t]
\includegraphics[height=.29\linewidth]{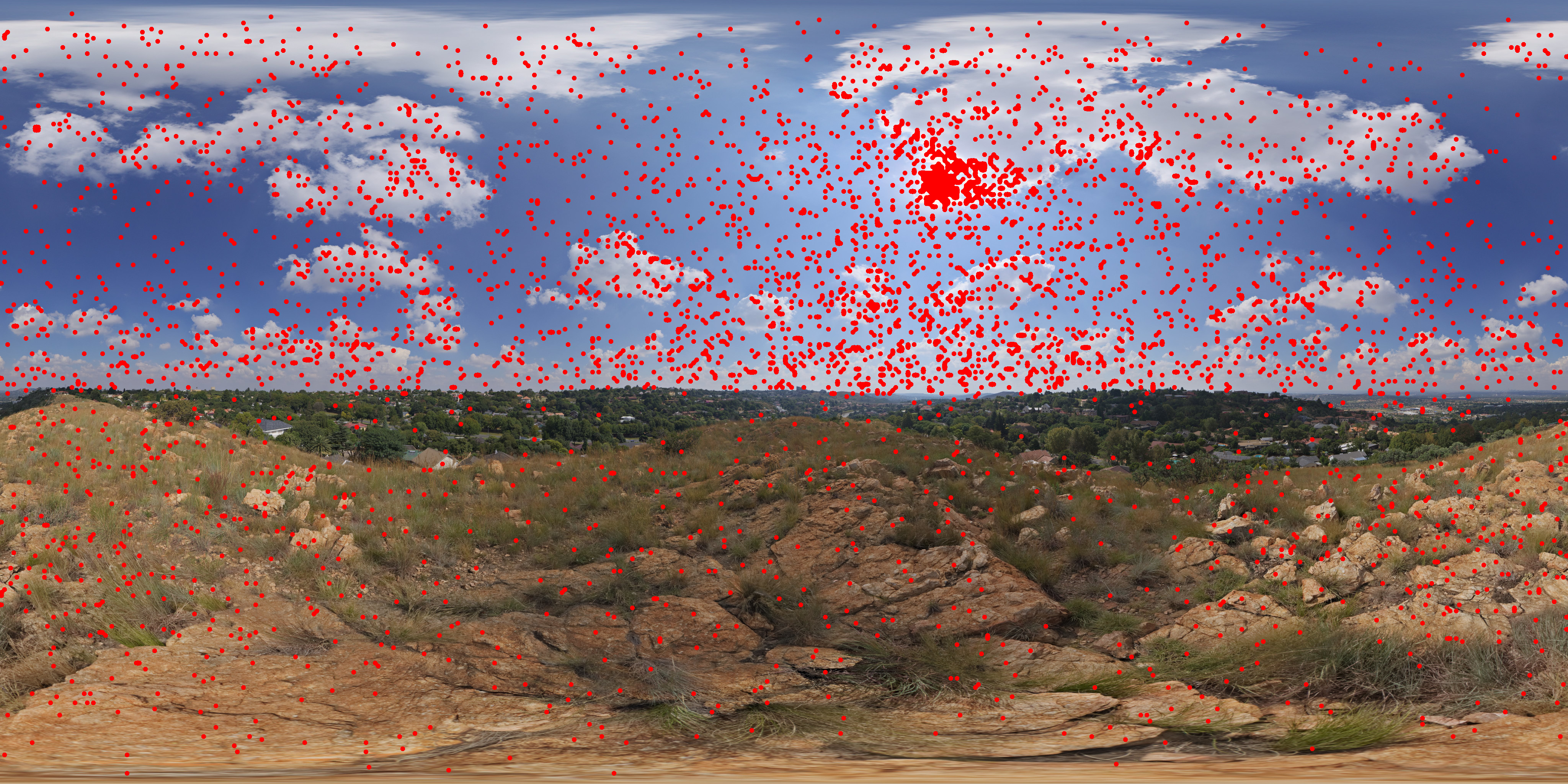}\hfill
\includegraphics[height=.29\linewidth]{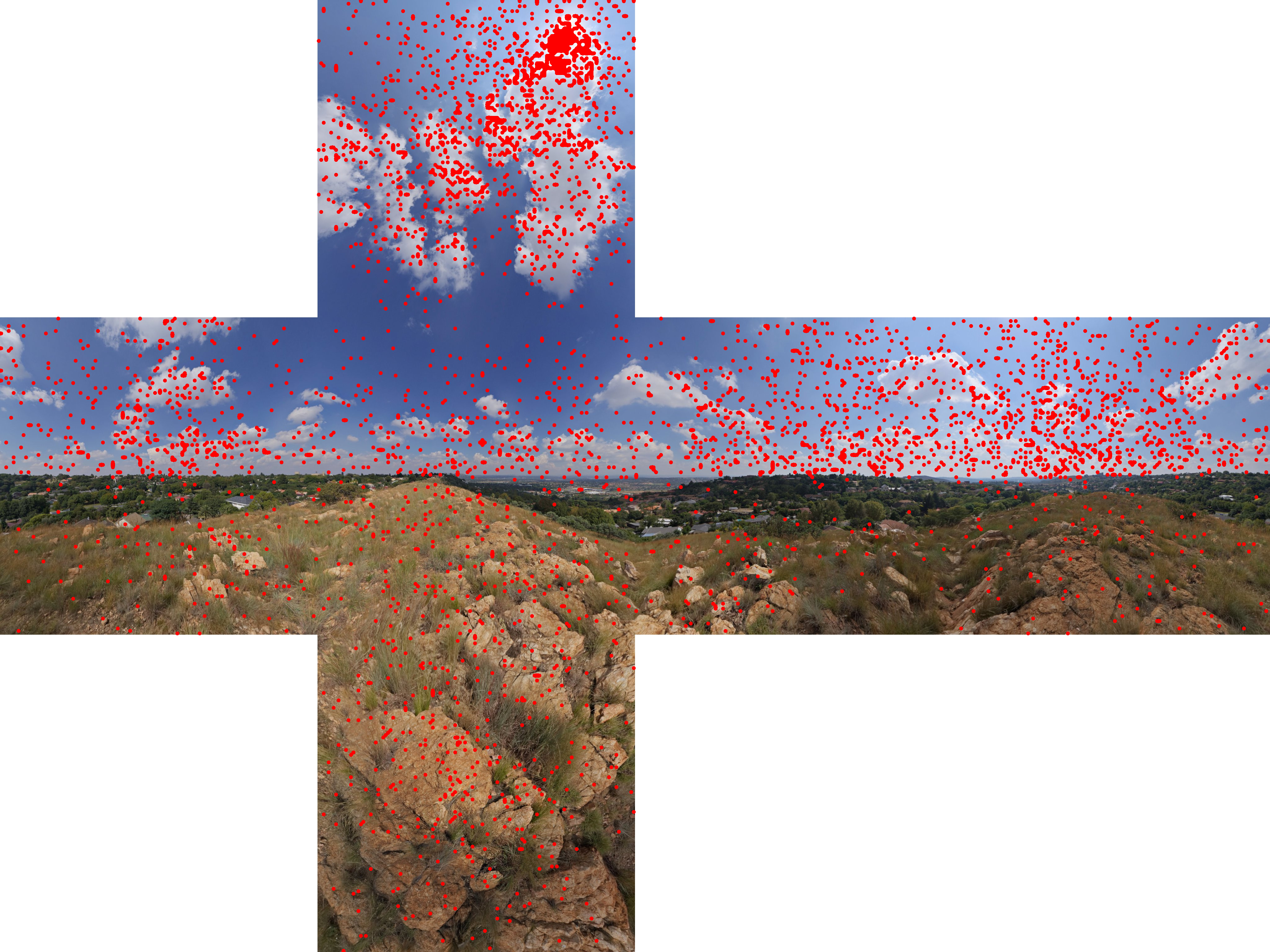}
\caption{\label{fig:samples-on-envmap}
5000 light direction samples taken with the method described in Sec.~\ref{sec:d}
overlaid over the equirectangular (left) and cube (right) environment map
parameterizations.}
\end{figure}

Fig.~\ref{fig:samples-on-envmap} shows 5000 light direction samples produced with
this method overlaid over the two parameterizations of the original
environment map.

\section{Results}

We implemented our importance sampling scheme as part of a Monte Carlo path
tracer with support for multiple importance sampling.

\begin{figure}[t]
\includegraphics[width=\linewidth]{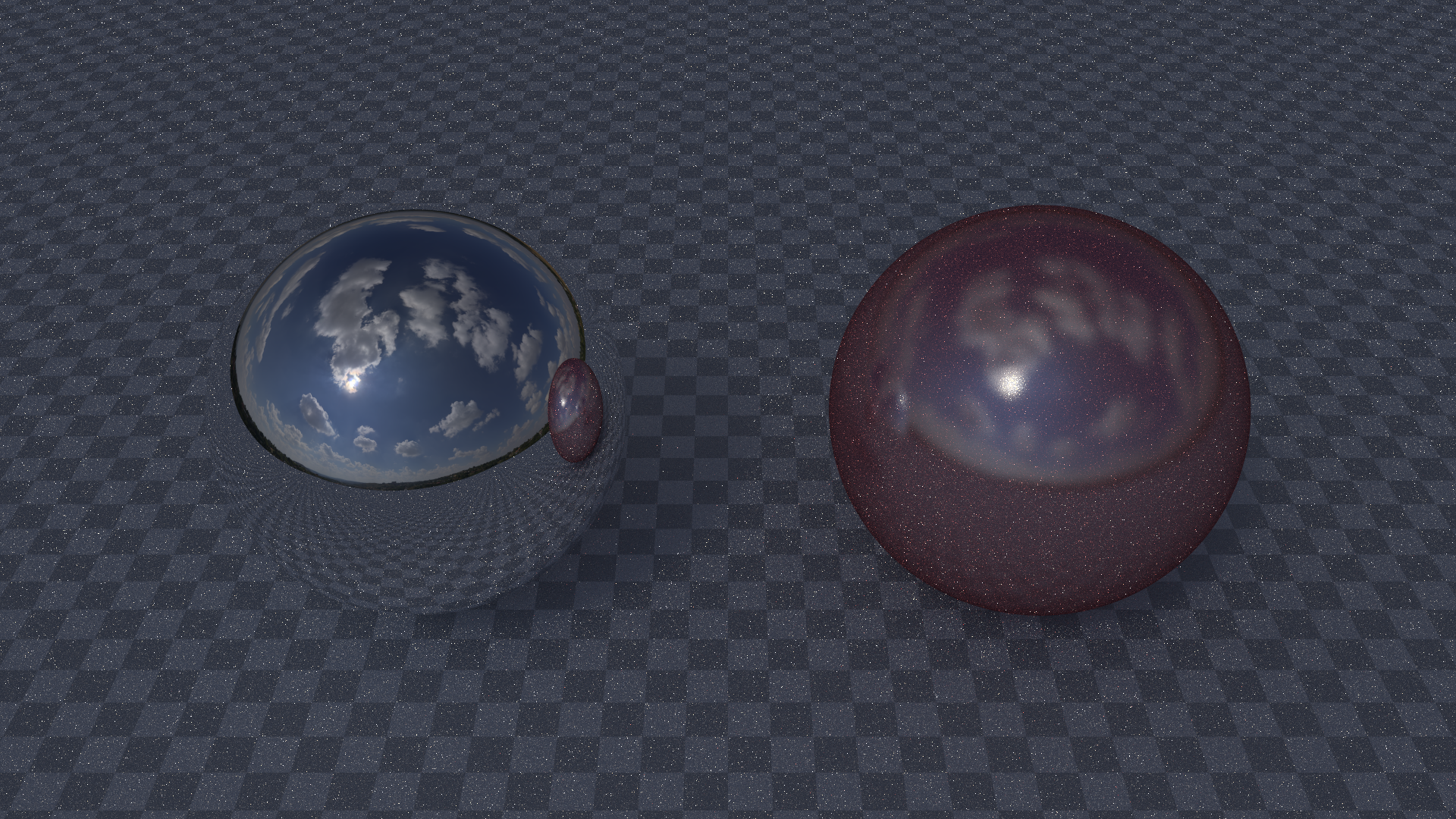}\\[1ex]
\includegraphics[width=\linewidth]{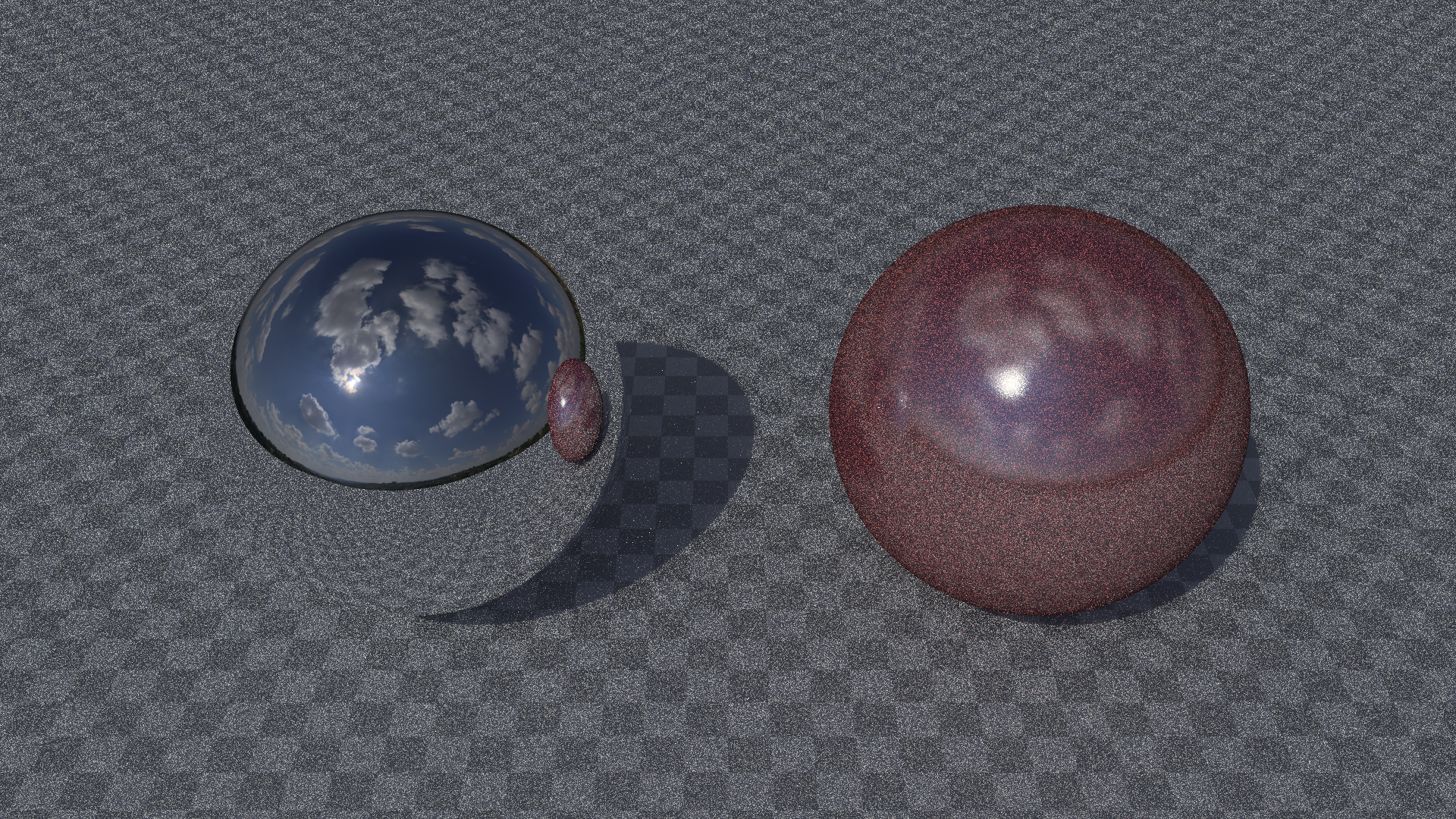}\\[1ex]
\includegraphics[width=\linewidth]{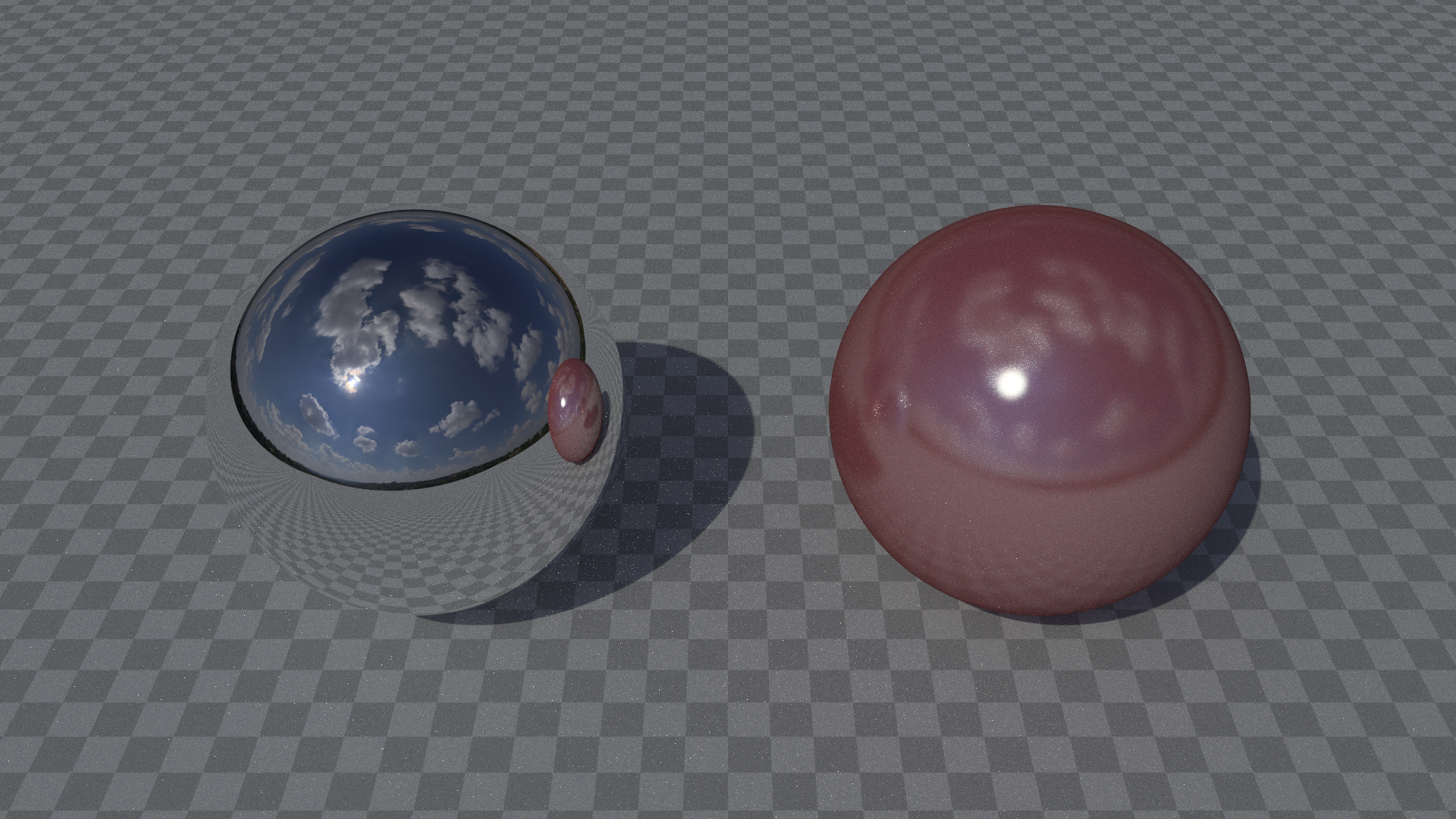}
\caption{\label{fig:results1}
A scene rendered at 64 samples per pixel without (top) and with environment map importance
sampling with parameter $N=256$ (middle) and $N=1024$ (bottom).}
\end{figure}

Fig.~\ref{fig:results1} shows an example scene rendered at 64 samples per pixel without and with environment map
importance sampling. The importance map size parameter $N$ was doubled in each
step. For values of $N$ greater than 1024, no quality improvements were visible,
which suggests that the underlying equirectangular environment map of size
$4096 \times 2048$ is adequately represented by an importance map of size
$1024\times 1024$. Note that both Mitsuba and pbrt build importance map structures at the
equirectangular image resolution instead.
The mirroring sphere on the left demonstrates that the resolution of the
environment map parameterization remains the same regardless of parameter $N$.

Rendering the same scene but with a cube environment map parameterization with a
resolution of $1024\times 1024$ per cube side does not result in noticeable
differences, as expected.

Modifying the scene by making the right sphere perfectly diffuse
enabled reproducing scene and parameters as closely as possible for the
Mitsuba renderer. For 64 samples per pixel and $N=1024$, there were no visible
quality differences to Mitsuba's result, confirming that
higher resolution importance map structures are not necessary for this typical
example of a daylight sky map.

\begin{figure}[t]
\includegraphics[width=\linewidth]{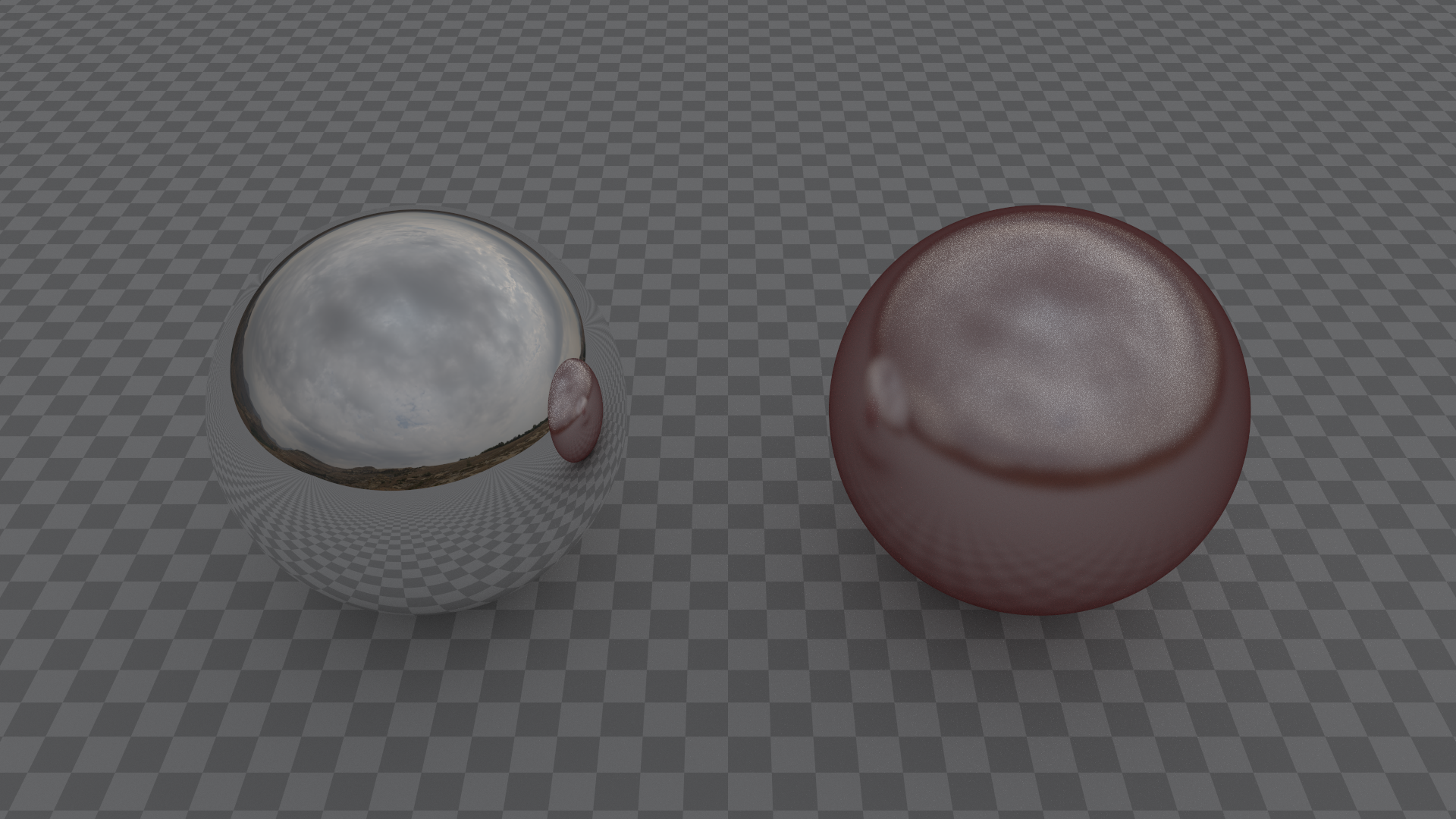}\\[1ex]
\includegraphics[width=\linewidth]{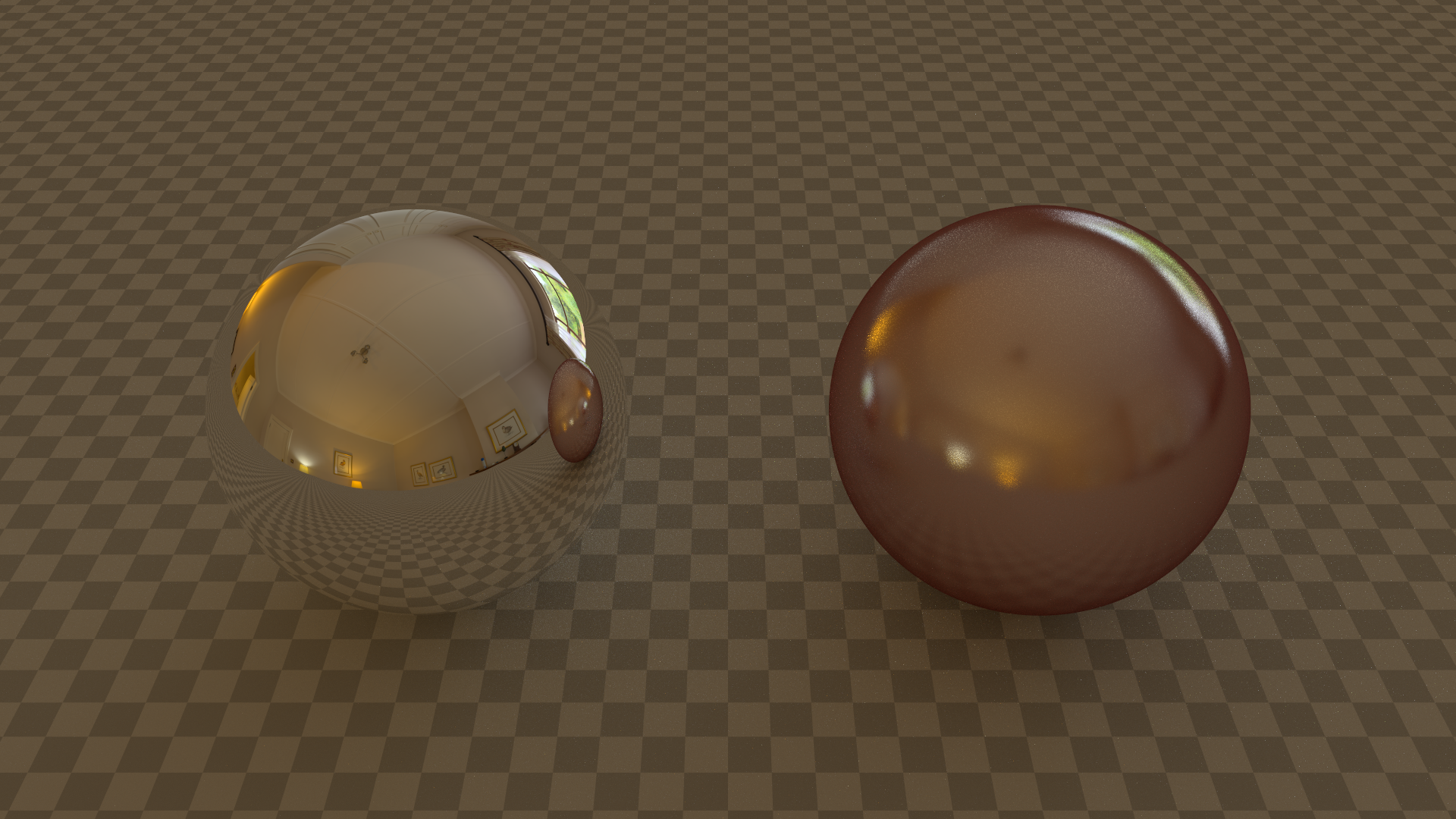}
\caption{\label{fig:results2}
A scene rendered with environment maps with more evenly distributed
lighting at 256 samples per pixel. For the cloudy sky map (top), $N=64$ is
sufficient for good quality, and $N=128$ for the indoor map (bottom).}
\end{figure}

Fig.~\ref{fig:results2} shows results for environment maps with more evenly
distributed lighting: a cloudy sky map and an indoor map, both given as
equirectangular maps of size $4096 \times 2048$. For such maps, $N$ can
be chosen significantly lower without loss of quality: $N=64$ for the cloudy sky
and $N=128$ for the indoor map. This results in significant reduction of memory
consumption.

\section{Conclusion}

This paper proposes an importance sampling scheme for environment maps that is
independent of their parameterization, so that it works with
equirectangular maps, cube maps, low distortion cube map
variants, or any other parameterization of the sphere surface.

Sampling the environment map is still performed on its original
parameterization so that sampling quality is unaffected and no resampling is
necessary.
The importance sampling scheme uses a separate map projection that is free of area
distortions to simplify the necessary computations. 

The granularity of importance sampling can be adapted to the
environment map contents to save memory. If this adaptation is undesirable,
for example if the contents of the environment map are not known in advance, the
parameter $N$ can always be chosen so that the importance map resolution is
similar to the resolution of the original environment map,
resulting in quality and memory consumption comparable to the
default choice of renderers such as Mitsuba and pbrt.

A limitation of the approach is that the bin size in the importance map is fixed
and must accommodate the smallest dominant light source in the environment,
resulting in bins that are smaller than necessary in homogeneous areas. This
could be alleviated by using a quadtree hierarchy instead of a fixed grid,
with a split criterion based on the variance of importance values inside a bin.
This hierarchical approach would also entirely remove the need to choose parameter $N$.

Note that the same methods can also be applied to data
given on a hemisphere, e.g.~measured BRDFs, by using the hemisphere variant of
the Lambert Azimuthal Equal-Area projection.

\section*{Acknowledgements}

The environment maps used in this paper are the high dynamic range maps ``Kloofendal Partly Cloudy'',
``Fouriesburg Mountain Cloudy'' and ``Lythwood Room'', provided under license
CC0 by \href{https://polyhaven.com/}{polyhaven.com}.

\section*{Supplementary Material}

The supplementary material contains C++ code that implements the importance
sampling scheme proposed in this paper in ca.~150 lines of code.

\newpage
\bibliographystyle{IEEEtran} 
\bibliography{paper.bib}

\begin{thebibliography}{10}
\providecommand{\url}[1]{#1}
\csname url@samestyle\endcsname
\providecommand{\newblock}{\relax}
\providecommand{\bibinfo}[2]{#2}
\providecommand{\BIBentrySTDinterwordspacing}{\spaceskip=0pt\relax}
\providecommand{\BIBentryALTinterwordstretchfactor}{4}
\providecommand{\BIBentryALTinterwordspacing}{\spaceskip=\fontdimen2\font plus
\BIBentryALTinterwordstretchfactor\fontdimen3\font minus
  \fontdimen4\font\relax}
\providecommand{\BIBforeignlanguage}[2]{{%
\expandafter\ifx\csname l@#1\endcsname\relax
\typeout{** WARNING: IEEEtran.bst: No hyphenation pattern has been}%
\typeout{** loaded for the language `#1'. Using the pattern for}%
\typeout{** the default language instead.}%
\else
\language=\csname l@#1\endcsname
\fi
#2}}
\providecommand{\BIBdecl}{\relax}
\BIBdecl

\bibitem{pharr2016pbr}
M.~Pharr, W.~Jakob, and G.~Humphreys, \emph{Physically based rendering: From
  theory to implementation}, 3rd~ed.\hskip 1em plus 0.5em minus 0.4em\relax
  Morgan Kaufmann, 2016.

\bibitem{greene1986cubemap}
N.~Greene, ``Environment mapping and other applications of world projections,''
  \emph{IEEE Comp. Graph. and Applications}, vol.~6, no.~11, pp. 21--29, Nov
  1986.

\bibitem{lambers2020cubemaps}
M.~Lambers, ``Survey of cube mapping methods in interactive computer
  graphics,'' \emph{The Visual Computer}, vol.~36, no.~5, pp. 1043--1051, May
  2020.

\bibitem{clarberg2008product}
P.~Clarberg and T.~Akenine-Möllery, ``Practical product importance sampling
  for direct illumination,'' \emph{Computer Graphics Forum}, vol.~27, no.~2,
  pp. 681--690, 2008.

\bibitem{estevez2018product}
A.~Conty~Estevez and P.~Lecocq, ``Fast product importance sampling of
  environment maps,'' in \emph{SIGGRAPH 2018 Talks}, 2018.

\bibitem{kollig2002diffuse}
T.~Kollig and A.~Keller, ``Efficient illumination by high dynamic range
  images,'' 2002.

\bibitem{brogers2014importance}
T.~Bashford-Rogers, K.~Debattista, and A.~Chalmers, ``Importance driven
  environment map sampling,'' \emph{IEEE Trans. Visualization and Computer
  Graphics}, vol.~20, no.~6, pp. 907--918, 2014.

\bibitem{bitterli2015portal}
B.~Bitterli, J.~Novák, and W.~Jarosz, ``Portal-masked environment map
  sampling,'' \emph{Computer Graphics Forum}, vol.~34, no.~4, pp. 13--19, 2015.

\bibitem{havran2005video}
V.~Havran, M.~Smyk, G.~Krawczyk, K.~Myszkowski, and H.-P. Seidel, ``Interactive
  system for dynamic scene lighting using captured video environment maps,'' in
  \emph{Eurographics Symp. Rendering}, K.~Bala and P.~Dutre, Eds., 2005.

\bibitem{agarwal2003structured}
S.~Agarwal, R.~Ramamoorthi, S.~Belongie, and H.~W. Jensen, ``Structured
  importance sampling of environment maps,'' in \emph{Proc. SIGGRAPH}, 2003, p.
  605–612.

\bibitem{lu2013realtime}
H.~Lu, R.~Pacanowski, and X.~Granier, ``Real-time importance sampling of
  dynamic environment maps,'' in \emph{Proc. Eurographics 2013 - Short Papers},
  May 2013, pp. 65--68.

\bibitem{snyder1987mapprojections}
J.~Snyder, \emph{Map projections---a working manual}, ser. Professional
  Paper.\hskip 1em plus 0.5em minus 0.4em\relax US Geological Survey, 1987,
  vol. 1395.

\bibitem{shirley1997map}
P.~Shirley and K.~Chiu, ``A low distortion map between disk and square,''
  \emph{Journal of Graphics Tools}, vol.~2, no.~3, pp. 45--52, 1997.

\bibitem{lambers2016mappings}
M.~Lambers, ``Mappings between sphere, disc, and square,'' \emph{Journal of
  Computer Graphics Techniques}, vol.~5, no.~2, pp. 1--21, Apr. 2016.

\end{thebibliography}

\end{document}